\documentclass[]{raa}                  
\usepackage{graphicx,times}             
\input{epsf.sty}                        
\input{psfig.sty}                       

\begin{document}

   \title{Magnetic Interaction in Ultra-compact Binary Systems
}

   \volnopage{Vol.0 (200x) No.0, 000--000}      
   \setcounter{page}{1}          

   \author{Kinwah Wu
   }

   \institute{ 
               Mullard Space Science Laboratory, University College London, Holmbury St.\ Mary,  
      Surrey RH5 6NT, United Kingdom  \\
   }

   \date{Received~~2009 month day; accepted~~2009~~month day}

\abstract{ 
  This article reviews the current works on ultra-compact double-degenerate binaries 
     in the presence of magnetic interaction, in particular, unipolar induction. 
  The orbital dynamics and evolution of compact white-dwarf pairs are discussed 
     in detail. 
  Models and predictions of electron cyclotron masers 
    from unipolar-inductor compact binaries and unipolar-inductor white-dwarf planetary systems 
    are presented. 
  Einstein-Laub effects in compact binaries are briefly discussed.   
}

   \authorrunning{K.~Wu}            
   \titlerunning{Magnetic interaction in ultra-compact binary systems}  

   \maketitle

%
%



\section{Introduction}           
\label{sect:intro} 

A binary could be a double-star system, star-planet system or planet-moon system  
  in which two objects revolve around each other under gravity.  
For binaries with a circular orbit,  
  the separation $a$ of the two components and the orbital period $P_{\rm o}$ 
  are related by  
\begin{eqnarray}  
 a & = &  \left[ G(M_1 + M_2)\left( \frac{P_{\rm o}}{2\pi} \right)^2\right]^{1/3}   \cr    
    & \approx  & 1.1\times 10^{10} \left(\frac{M_1+M_2}{{\rm M}_\odot} \right)^{1/3} 
         \left(\frac{P_{\rm o}}{600~{\rm s}} \right)^{2/3} \ {\rm cm}\ , 
\label{eq:kepler}
\end{eqnarray}  
 where $M_1$ and $M_2$ are the masses of the primary and secondary components respectively, 
  and $G$ is the gravitational constant.   
(Hereafter the subscripts ``o'', ``1" and ``2" represent 
   the orbit, the primary star and the secondary star respectively.) 
If the secondary component fills its Roche lobe, 
  its mean density ${\bar \rho}_2$ is determined by the orbital period:      
\begin{eqnarray} 
   {\bar \rho}_2  & \approx &   3.9 \times 10^{3}\   \lambda(q) 
      \left(\frac{P_{\rm o}}{600~{\rm s}} \right)^{-2}~{\rm g}~{\rm cm}^{-3} \  ,     
\label{period_density}
\end{eqnarray}   
   where $\lambda(q)$ is a numerical factor of the order of unity, 
   weakly dependent on the mass ratio $q = M_2/M_1$ 
   (Eggleton 1983; see also Pacyznski 1971).   
In a stellar binary with $P_{\rm o} \approx 600~{\rm s}$,  
  the density of the secondary star would exceed that of a main-sequence star with the same mass.  
The primary star, which is more massive, is even denser. 
Thus, stellar binaries with $P_{\rm o} < 600~{\rm s}$ 
   must contain either degenerate stars or black holes, 
   and these short-period systems are known as ultra-compact double-degenerate systems (UCDs). 

In principle, UCDs may contain any combinations of white dwarfs, neutron stars or black holes. 
However, the formation of double white dwarfs are more favourable in the evolutionary channels 
  (see Han 1998; Nelemans et al.\ 2001; Belczynski \& Taam 2004; 
  Postnov \& Yungelson 2006; Belczynski et al.\ 2008), 
  and compact double white dwarfs are expected to be more abundant 
  than compact binaries with other combinations of white dwarfs, neutron stars and black holes.  
Observationally, many compact white-dwarf pairs have been discovered 
  (see Roelof, Nelemans \& Groot 2007), 
  and these system are populous in the Milky Way.  
In this article the main focus will be on double white-dwarf systems.  
Hereafter, unless otherwise stated, 
 the term UCD will be used solely for short-period systems with two white dwarfs.  
  
Almost all celestial bodies possess a certain magnetism. 
A substantial fraction of white dwarfs are known to have a magnetic field 
   with strength exceeding $10^6$~G  
   (Chanmugam 1992; Schmidt \& Smith 1995; Wickramasinghe \& Ferrario 2000). 
The magnetic moments of these white dwarfs are above $10^{32}$~G~cm$^3$. 
For an orbital separation $< 10^{10}$~cm,  
  these white dwarfs will exert a magnetic field of the order of kG 
  at the surface of their companion stars.  
As the two white dwarfs in UCDs are in very close proximity,  
  electromagnetic interaction is inevitable. 
This alters the orbital dynamics of the binary 
  and gives rise to a variety of unusual observational consequences.   

Magnetic interaction between two gravitationally bound celestial objects 
  is common on all scales.   
A well known example in our backyard is Jupiter and its moon Io.   
It is believed that Io has a highly conductive core. 
When Io revolves around Jupiter, it traverses the Jovian magnetic field 
  and a large e.m.f. is created via a unipolar-induction process 
  (Piddington \& Drake 1968; Goldreich \& Lynden-Bell 1969). 
This e.m.f. drives the flow of electric currents between Jupiter and Io.  
Observations have shown a hot spot at the polar surface of Jupiter (Clarke et al.\ 1996),  
  which is identified as the location of foot-points of the magnetic field lines 
  leading to Io.  
Dissipation of the electric currents in the Jovian atmosphere 
  lights up the foot-points of the magnetic field lines that connect the two objects.    
On stellar scales, strong magnetic interactions are found 
  between the two stars in RS CVn binaries and in AM Herculis binaries. 
There is also evidence that substantial magnetic interaction occurs in Algol binaries as well  
   (Richards \& Albright 1993; Retter, Richards \& Wu 2005). 
In RS CVn binaries the magnetic interaction 
  leads to enhanced coronal activity in the component stars 
  (Uchida \& Sakurai 1983; Ferreira \& Mendoza-Brice\~no 2005). 
In AM Herculis binaries, 
  magnetic interaction essentially defines the characteristics of the system.  
It locks the entire system to into synchronous rotation 
  (Campbell 1983, 1999; Wickramasinghe \& Wu 1991; Wu \& Wickramasinghe 1993); 
  it governs their orbital evolution (Li, Wu \& Wickramasinghe 1994a, b; Davis et al.\ 2008); 
  and it determines the hydrodynamics of mass flow 
  from the Roche-lobe spilling low-mass donor star to the magnetic white dwarf primary 
  (Chanmugam \& Wagner 1977; Visvanathan \& Wickramasinghe 1981, 
   see also Warner 1995; Wu 2000).   
  
It is natural that the white dwarfs in UCDs interact magnetically, 
  provided that one or two of the white dwarfs have a sufficiently large magnetic moment.   
In this article we will review the current research progress on magnetically interacting UCDs  
  and associated systems. 
We organise the article as follows. 
In \S2 we discuss the general orbital dynamics of UCDs in compact binaries. 
In \S3 we present the basics of the unipolar induction model for compact white-dwarf pairs,  
  and in \S4 we discuss the orbital evolution of compact binaries 
    in the presence of unipolar induction.  
In \S5 we show that unipolar-inductor white-dwarf pairs could be  
   electron-cyclotron maser sources. 
In \S6 we show how some physics in UCDs  
  can be applied to related systems, such as white dwarf-planet systems, 
  and that magnetically interacting ultra-compact binaries 
    may exhibit Einstein-Laub effects.  
 
 
\section{Orbital dynamics in compact binaries} 
\label{sect:dynamics}

AM CVn binaries are the better studied UCD (Solheim 1995; Nelesmans 2005). 
Mass transfer occurs in AM CVn binaries 
  when the less massive white dwarf overfills its Roche lobe. 
The in-falling material forms an accretion disk around the white-dwarf primary. 
If the binary orbit is too compact, the formation of an accretion disk might be prohibited. 
Mass transfers directly via a gas stream 
  from the inner Lagrangian point of the secondary white dwarf 
  to the surface of the primary white dwarf (Marsh \& Steeghs 2002, see also Wood 2009). 
The mass transfer dynamics of these double white dwarfs are analogous to 
  those of the Algol binaries. 
These binaries are known as direct-impact mass-transfer double degenerates.  
The orbital dynamics and evolution 
  of AM CVn binaries and direct-impact mass-transfer double degenerates 
    are regulated by the mass transfer process. 
Their high-energy emissions, such as X-rays, are accretion powered.  
Magnetic interacting UCDs are similar to AM CVn binaries and 
  direct-impact mass-transfer double degenerates,  
  as they also have two white dwarfs revolving around each other in a very tight orbit. 
However, they are different from those binaries 
  in that magnetic interaction governs the angular momentum redistribution,  
  and that internal energy dissipation within the system 
  plays an important role in regulating the orbital dynamics 
  and hence the orbital evolution.    
  
  
\begin{figure}
   \vspace{0.05mm}
   \begin{center}
   \hspace{3mm}\psfig{figure=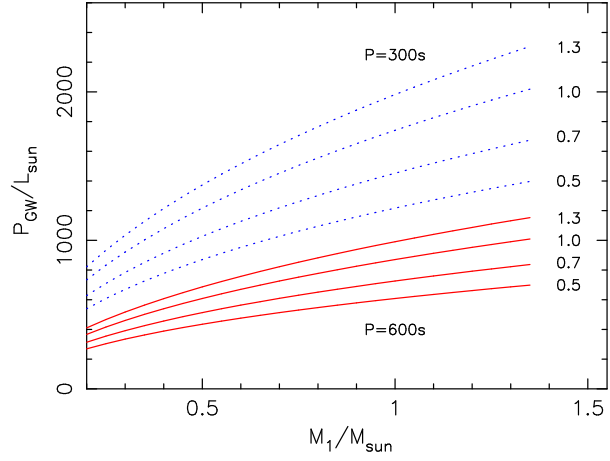,width=90mm,angle=0.0}
   \caption{
     The power of gravitational waves from white-dwarf pairs, 
          normalised to the solar bolometric luminosity, 
          as a function of the primary white-dwarf mass $M_1$,  
          for orbital periods of 600~s (solid lines) and 300~s (dotted lines). 
      Each curve corresponds to a value of the secondary white-dwarf mass, 
          labeled in solar-mass unit. (Adapted from Wu, Ramsay \& Willes (2008).) 
               } 
\label{gw_power}               
   \end{center}
\end{figure} 


UCDs are strong sources of gravitational waves because of their compact orbits.  
The power of their gravitational radiation (assuming an orbital eccentricity $e= 0$) is 
\begin{eqnarray}  
   {\dot E}_{\rm gw} & = & -\  \frac{32}{5} \frac{G^4}{c^5} 
      \frac{M_1^2 M_2^2 \left(M_1+M_2 \right)}{a^5}  \nonumber \\  
        & = & -\  \frac{32}{5} \frac{G^{7/3}}{c^5} 
        \ M_{\rm chirp}^{10/3}\  \omega_{\rm o}^{10/3} \nonumber \\  
     & = & -\ 1.2 \times 10^{36} \left[ 
       \left(   \frac{M_{\rm chirp}}{{\rm M}_\odot} \right)  
        \left(  \frac{600~{\rm s}}{P_{\rm o}} \right) \right]^{10/3} ~{\rm erg}~{\rm s}^{-1} 
\label{eq:gw_power}
\end{eqnarray} 
  (see Landau \& Lifshitz 2002),   where $c$ is the speed of light. 
The chirp mass $M_{\rm chirp} = {\tilde M}^{3/5}(M_1+M_2)^{2/5}$, 
  where ${\tilde M}  \equiv M_1 M_2/(M_1+ M_2)$ is the reduced mass of the binary.  
It relates the orbital angular momentum $J_{\rm o}$ 
  to the orbital angular velocity $\omega_{\rm o}$ via 
\begin{eqnarray}   
  J_{\rm o} & = &  G^{2/3} M_{\rm chirp}^{5/3} \omega_{\rm o}^{-1/3} \ . 
\label{ang_mom}
\end{eqnarray}    
For  UCDs with $P_{\rm o} \sim 600~{\rm s}$ or shorter, 
   the power of the gravitational radiation 
   greatly exceeds the solar power in the electromagnetic spectrum (Fig.~\ref{gw_power}).   
  
The orbital angular momentum of a binary system is given by 
$J_{\rm o}  =  M_1 M_2~a^2 \omega_{\rm o} / (M_1 + M_2)$.     
The orbital separation $a$ and the  orbital angular velocity $\omega_{\rm o}$ 
  are related by 
$ \omega_{\rm o}^2   = (2\pi/P_{\rm o})^2=  G~ (M_1+M_2)~a^{-3}$. 
The evolution of the binary orbit 
  is determined by redistribution of angular momentum within the system 
  and the loss of angular momentum from the system.  
These processes are described by the following coupled differential equations:  
\begin{eqnarray} 
  \frac{\dot{\omega}_{\rm o}}{\omega_{\rm o}} & = 
     & \left( 3 - \frac{M_1}{M_1+M_2}\right) {\frac{\dot M_1}{M_1}} 
        + \left( 3 - \frac{M_2}{M_1+M_2}\right) {\frac{\dot M_2}{M_2}} 
        - 3 \frac{\dot J_{\rm o}}{J_{\rm o}} \ , 
\label{eq:evolution_1} \\ 
   \frac{\dot{\omega}_1}{\omega_1} & = 
       & - \left(1+ \frac{2}{n_1} \right) \frac{\dot M_1}{M_1} + \frac{\dot J_1}{J_1} \ , 
\label{eq:evolution_2} \\ 
 \frac{\dot{\omega}_2}{\omega_2} & = 
      & - \left(1+ \frac{2}{n_2} \right) \frac{\dot M_2}{M_2} + \frac{\dot J_2}{J_2} \  ,  
\label{eq:evolution_3}
\end{eqnarray}  
    where  ``~$\cdot$~'' denotes time derivatives.   
The derivation of the above equations has assumed that 
 $M_1  \propto R_1^{n_1}$ and $M_2  \propto R_2^{n_2}$,  
    where $n_{1,2}$ are the proportional indices in the mass-radius relations of the two stars. 

Conservation of angular-momentum requires 
${\dot J} = {\dot J}_{\rm o} + {\dot J}_1+ {\dot J}_2$. 
When there is no mass loss from the system (${\dot M} ={\dot M}_1 + {\dot M}_2 = 0$), 
   orbital angular momentum is extracted from the binary 
   only through the emission of gravitational waves.  
This gives a rate of orbital angular-momentum loss 
\begin{eqnarray} 
  {\dot J} & = & {\dot J}_{\rm gw} \nonumber \\ 
               & = &  -\ \frac{32}{5} \frac{G^{7/2}}{c^5}
                    \frac{M_1^2 M_2^2 (M_1+M_2)^{1/2}}{a^{7/2}} \nonumber  \\ 
                        & = & -\  \frac{32}{5} \frac{G^{7/3}}{c^5} 
        \ M_{\rm chirp}^{10/3}\  \omega_{\rm o}^{7/3} 
\label{eq:j_dot}        
\end{eqnarray}  
      (Landau \& Lifshitz 2002).    
If there is no mass change between the two stars 
   (${\dot M}_1 = {\dot M}_2 = {\dot M}_{\rm chirp} = 0$) 
   and if the stellar spins are decoupled from the orbital rotation,    
  the evolution of the orbital angular frequency is dictated by gravitational radiation loss: 
\begin{eqnarray}  
   \frac{{\dot \omega}_{\rm o}}{\omega_{\rm o}} 
      & = & - \ 3 \frac{{\dot J}_{\rm gw}}{J_{\rm o}} \nonumber \\ 
      & = & \frac{96}{5} \frac{G^{5/3}}{c^5} M_{\rm chirp}^{5/3} \omega_{\rm o}^{8/3} \  .  
\label{eq:orbit_1}
\end{eqnarray}   
It is clear that in the absence of mass loss from the system and 
  in the absence of angular momentum exchange or mass exchange between the two stars,  
  the binary orbit is always spun up, i.e.\ the orbital period decreases with time.   

If the stellar spins are coupled with the orbital rotation, 
  then angular momenta can be injected from the orbit into the stars. 
In the `ideal' case where the two stars and the orbit  
  are locked in synchronous rotation,  
\begin{eqnarray}  
   \frac{{\dot \omega}_{\rm o}}{\omega_{\rm o}} 
      & = & - \ 3 \frac{{\dot J}_{\rm gw}}{J_{\rm o}}  
        \left[ 1- \frac{3}{J_{\rm o}} \left(J_1+J_2\right) \right]^{-1}    \nonumber \\ 
       & = & \frac{96}{5} \frac{G^{5/3}}{c^5} M_{\rm chirp}^{5/3} \omega_{\rm o}^{8/3}  
         \left[ 1- \frac{3(M_1 R_1^2 + M_2 R_2^2) \omega_{\rm o}^{4/3}}
        {G^{2/3} M_{\rm chirp}^{5/3}}   \right]^{-1}    \  .   
\label{eq:orbit_2}
\end{eqnarray}    
Internal energy dissipation in the system is unimportant 
   in an ideal, perfectly synchronous rotating system.   
When $M_{\rm chirp}$ is fixed, $\omega_{\rm o} \propto J_{\rm o}^3$. 
As additional angular momentum is extracted from the orbit to spin up the two stars, 
   the orbital angular frequency will accelerate further 
   when the system loses energy via gravitational radiation.  
This gives larger values for ${\dot \omega}_{\rm o}$ than those in the case 
   where the spins of the star and the orbital rotation are decoupled 
   (cf.\ Eq.~\ref{eq:orbit_1} and \ref{eq:orbit_2}).  

In reality, perfect synchronism is hard to achieve for any binary system. 
Although AM Herculis binaries are supposed to be magnetically locked 
   into synchronous rotation, 
   there are a small fraction (e.g.\ the system BY Cam, Mason et al.\ 1998) 
   in which the white dwarf rotates asynchronously with the orbital motion.  
The situation is similar for UCDs. 
There should be certain spin-orbit asynchronism 
   despite the fact that strong tidal force and magnetic interaction 
   tend to synchronise the star spins and orbital rotation.   
When there are internal energy dissipation and angular momentum redistribution in the system,     
   the formulation for the orbital evolutionary dynamics described above would need modifying.     
In the next section we will discuss the case of slightly asynchronous UCDs  
   in which magnetic interaction mediates  
   the angular momentum exchange between the stars and the orbit. 
Also, there is no mass transfer between the stars in these systems, 
   contrary to the magnetically locked AM Herculis binaries. 
The dynamics would be more complicated 
   when mass exchange occurs, and when the system loses mass.  
(Orbital evolution of binaries under mass exchange and mass outflow 
   were discussed, for example, in Wu 1997.)   
   

\section{Unipolar induction in compact binaries}  
\label{sect:unipolar_induction}
   
The small separation between the stars in a UCD 
   allows electromagnetic interactions to occur between them. 
One possible process is unipolar induction, 
  which could generate strong electric currents between the two white dwarfs,  
  as well as  large Lorentz torques on the orbit and the stars. 

Unipolar induction is a fundamental electrodynamic process.  
It is a manifestation of Maxwell's equations and the Lorentz force acting on electrons 
   (Feynman, Leighton \& Sands 1964; Assis 2000).  
Its validity is verified by laboratory experiments (see Miller 1981; Kelley 1999).  
A proper interpretation of unipolar induction is still under discussion, 
  as there are subtleties in how it is related to electrodynamics and relativity 
  (see, for example, recent articles by Montgomery 1999; 
   Guala-Valverde, Mazzoni \& Achilles 2002).    
A well known example of astrophysical unipolar inductors is the Jupiter-Io system  
   (Piddington \& Drake 1968; Goldreich \& Lynden-Bell 1969).   
It has been proposed that unipolar induction operates 
  in pulsar magnetospheres (Goldreich \& Julian 1969),  
  in magnetic binary stars (e.g.\ AM Herculis  binaries, Chanmugam \& Dulk 1982),  
  in stellar-planetary systems (see e.g.\ Zarka 2007; Laine, Lin \& Dong 2008), 
  in white-dwarf planetary systems (Li, Ferrario \& Wickramasinghe 1998; Willes \& Wu 2004), 
  and in magnetised accretion disks around black holes 
  (Shatskii 2003, see also Punsly 2001; Komissarov 2004).    
There are also models wherein cosmic ray particles are accelerated to ultra-high energies  
  via unipolar induction (Chanmugam \& Brecher 1985; Shatskii \& Karashev 2002; 
  see also discussions in Blandford 2000).   
A unipolar-inductor model (sometimes known as electric-star model) for UCD was proposed
  (Wu et al.\ 2002; Ramsay et al.\ 2002; Willes, Wu \& Kuncic 2004; 
  Dall'Osso, Israel \& Stella 2006, 2007; Wu, Ramsay \& Willes 2008) 
  to explain the peculiar properties of the X-ray sources  RX~J1914+24 and RX~J0806+15.     
  

\begin{figure}
   \vspace{2mm}
   \begin{center}
   \hspace{3mm}\psfig{figure=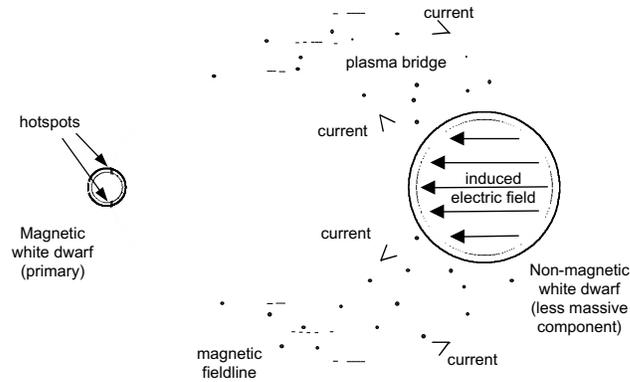,angle=0.0}
   \caption{A schematic illustration of the unipolar-inductor model for white-dwarf pairs. 
      As the system revolves, a large e.m.f.\  is induced across the non-magnetic white dwarf 
        and currents are driven between the two stars. 
      The resistance in the atmospheric layers of the white dwarfs causes energy dissipation. 
      Electromagnetic waves are emitted from the heated white-dwarf atmosphere.   }
   \label{ui_illustration}
   \end{center}
\end{figure} 


\subsection{Compact white-dwarf pairs} 

When a non-magnetic conducting body of linear size $R$     
   traverses a magnetic field $\vec B$  with a velocity $\vec v$,  
   an e.m.f.\  $\Phi  \sim   R | {\vec E} |$ is induced across the conducting body,     
   where ${\vec E} = {\vec \beta} \times {\vec B}$  
   and  ${\vec \beta} = {\vec v}/c$.    
This is the basic principle of the operation of unipolar-induction 
   in magnetically interacting white-dwarf pairs. 
The setting for a unipolar inductor UCD is illustrated in Figure~\ref{ui_illustration}.   
The e.m.f.\ across the non-magnetic white dwarf in orbit with a magnetic white dwarf  
   is therefore  
\begin{eqnarray}  
   \Phi  & \approx & {{2\pi} \over c} \biggl(  
      {{\mu_1 R_2} \over {a^2 P_{\rm o}}} \biggr) (1 - \alpha) \nonumber  \\ 
    &  = &  \biggl({{\mu_1 R_2} \over c} \biggr)  
        \biggl({{2\pi} \over P_{\rm o}}  \biggr)^{7/3} 
       (1 - \alpha) \big[G M_1 (1+q) \big]^{-2/3} \  , 
\label{eq:emf}
\end{eqnarray}   
   where $q$ ($\equiv M_2/M_1$) is the mass ratio  
   of the non-magnetic to the magnetic white dwarf, 
   $R_2$ and $R_1$ are the respective radii of the two white dwarfs, 
   and  $\mu_1$ is the magnetic moment of the magnetic white dwarf. 
The induced e.m.f.\ depends on the degree of synchronism 
  between the spin of the magnetic white dwarf and the orbit.  
Without loss of generality we may specify the degree of asynchronism 
  using a parameter $\alpha$, 
   which is the ratio of the spin angular speed of the magnetic white dwarf $\omega_1$   
   to the orbital angular speed  $\omega_{\rm o}$. 
(We consider the convention in which the anti-clockwise direction is positive.)  

Provided that the space between the white dwarfs is permeated by some plasma, 
   the e.m.f.\ will drive electric currents, 
   which flow along the magnetic field lines connecting the two white dwarfs.   
Although white dwarfs have a highly conducting core, 
   there is substantial electric resistance in the white-dwarf atmosphere, 
   where electrical dissipation occurs.  
The total power generated by the current dissipation in the two stars is  
\begin{eqnarray}   
   W &  = & I^2 ({\cal R}_1+{\cal R}_2) \nonumber  \\ 
     &  = & \frac{\Phi^2}{{\cal R}_1+ {\cal R}_2} \ ,    
\label{eq:diss}
\end{eqnarray} 
   where $I$ is the total current,   
   and ${\cal R}_1$ and ${\cal R}_2$ are the effective resistance 
   of the magnetic and the non-magnetic white dwarf respectively. 
For an object with a length $L$ and a cross-sectional area $A$,  
  the resistance is simply ${\cal R} = L /A\sigma$  (with $\sigma$ as electric conductivity). 
It follows that the ratio of the effective resistances of the white dwarfs is    
\begin{eqnarray} 
   {{\cal R}_1} \over {{\cal R}_2} & \sim & 
      \biggl({{\sigma_2}\over {\sigma_1}} \biggr)
       \biggl({{R_2^2}\over {fR_1^2}}\biggr)
       \biggl({{\Delta h_1}\over {\Delta h_2}}\biggr)   \ ,  
\label{eq:resistance}
\end{eqnarray}  
   where  $\sigma_1$ and $\sigma_2$ are the corresponding electric conductivities  
   of the two white dwarfs,  $\Delta h_1$ and $\Delta h_2$ are the thicknesses  
   of the dissipative surface layers of the white dwarfs, and 
   $f$ is the fractional effective area of the magnetic poles (hot spots) 
   on the surface of the magnetic white dwarf.    
As $f \ll 1$ (see Wu et al.\ 2002), 
   the effective resistance of the magnetic white dwarf
   is significantly larger than that of the non-magnetic white dwarf.   

As the electric currents pass through both white dwarfs,   
   the ratio of the power dissipation in the magnetic primary  
   to that of the non-magnetic secondary is 
   ${{W_1}/{W_2}} =  {{\cal R}_1} / {{\cal R}_2}$.  
Taking account of the geometry of the current loops,  we obtain 
\begin{eqnarray} 
    {{W_1} \over {W_2}}&  \approx & \zeta
      \biggl({{\sigma_2} \over {\sigma_1}} \biggr)   
      \biggl({{R_2} \over {\Delta R_2}} \biggr)     
      \biggl[{{G(M_1+M_2)} \over {R_1^3}}   
           \biggl({P_{\rm o} \over {2\pi}} \biggr)^2 \biggr]^{1/2}  \ ,  
\label{eq:power_ratio}   \\    
   {\cal R}_1 & \approx & {1\over {2 \sigma_1}} 
                    \biggl({{H}\over {\Delta d}}\biggr) 
                    \biggl({a \over R_1}\biggr)^{3/2} 
                    {{{\cal J}(e)} \over R_2} \ ,  
\label{eq:ris_1}                     \\ 
   {\cal R}_2 & \approx & {4\over {\pi \sigma_2}}  
                      \biggl({{\Delta R_2}\over {R_2^2}}\biggr)    
\label{eq:ris_2}
\end{eqnarray}  
  (see Appendices A and B of Wu et al.\ (2002) for details), 
   where $\Delta R_2$ is the thickness of the secondary's atmosphere 
   and $\zeta$ is a structure factor of the order of unity.  
The factor ${\cal J}(e)$ depends on the radii of the white dwarfs relative to the orbital separation. 
Its value is of the order of unity for white-dwarf pairs with $P_{\rm o}$ less than an hour.   

The electric conductivity of plasma at an electron temperature $T_{\rm e}$ is given by 
\begin{eqnarray} 
   \sigma & = & \gamma \biggl({{2^{5/2}}\over {\pi^{3/2}}} \biggr)
        {{(kT_{\rm e})^{3/2}}\over{m_{\rm e}^{1/2}Ze^2 \ln \Lambda}} \ , 
\label{eq:cond}
\end{eqnarray}    
   (Spitzer \& H$\ddot {\rm a}$rm 1953)  
   where $k$ is the Boltzmann constant, 
   $m_{\rm e}$ is the electron mass,  
   $e$ is the electron charge,  
   $Z$ is the ion charge number, 
   and $\ln \Lambda$ is the Coulomb logarithm. 
The factor $\gamma$ depends on $Z$, 
   which has values between 0.6 ($Z=1$) and 1 ($Z\rightarrow \infty$)  
   (see Alfv$\acute {\rm e}$n \& F$\ddot {\rm a}$lthammar 1963).    
For a white-dwarf atmosphere with $T_{\rm e} \sim 10^5$~K, 
   the conductivity $\sigma \sim 10^{13}-10^{14}$~esu. 
Since the conductivities of the atmospheres of the white dwarfs are similar to each other,   
   the majority of the electrical power will be dissipated 
   in small regions at the footpoints of the current-carrying field lines
   on the surface of the magnetic white dwarf.  

\begin{figure}
  \vspace*{0.25cm} 
 \begin{center}   
    \mbox{\epsfxsize=0.42\textwidth\epsfbox{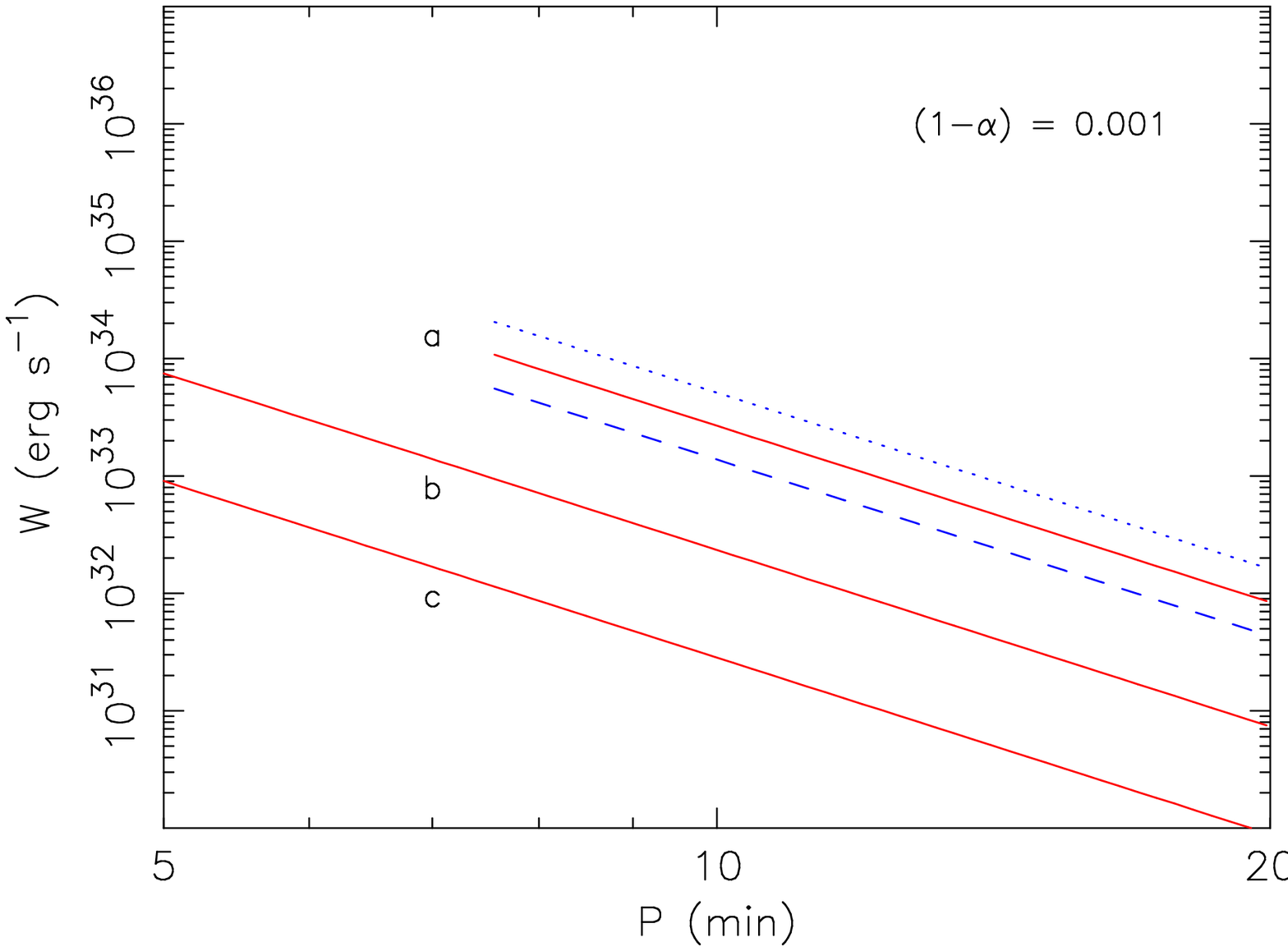}   \hspace*{0.2cm}
     \epsfxsize=0.42\textwidth\epsfbox{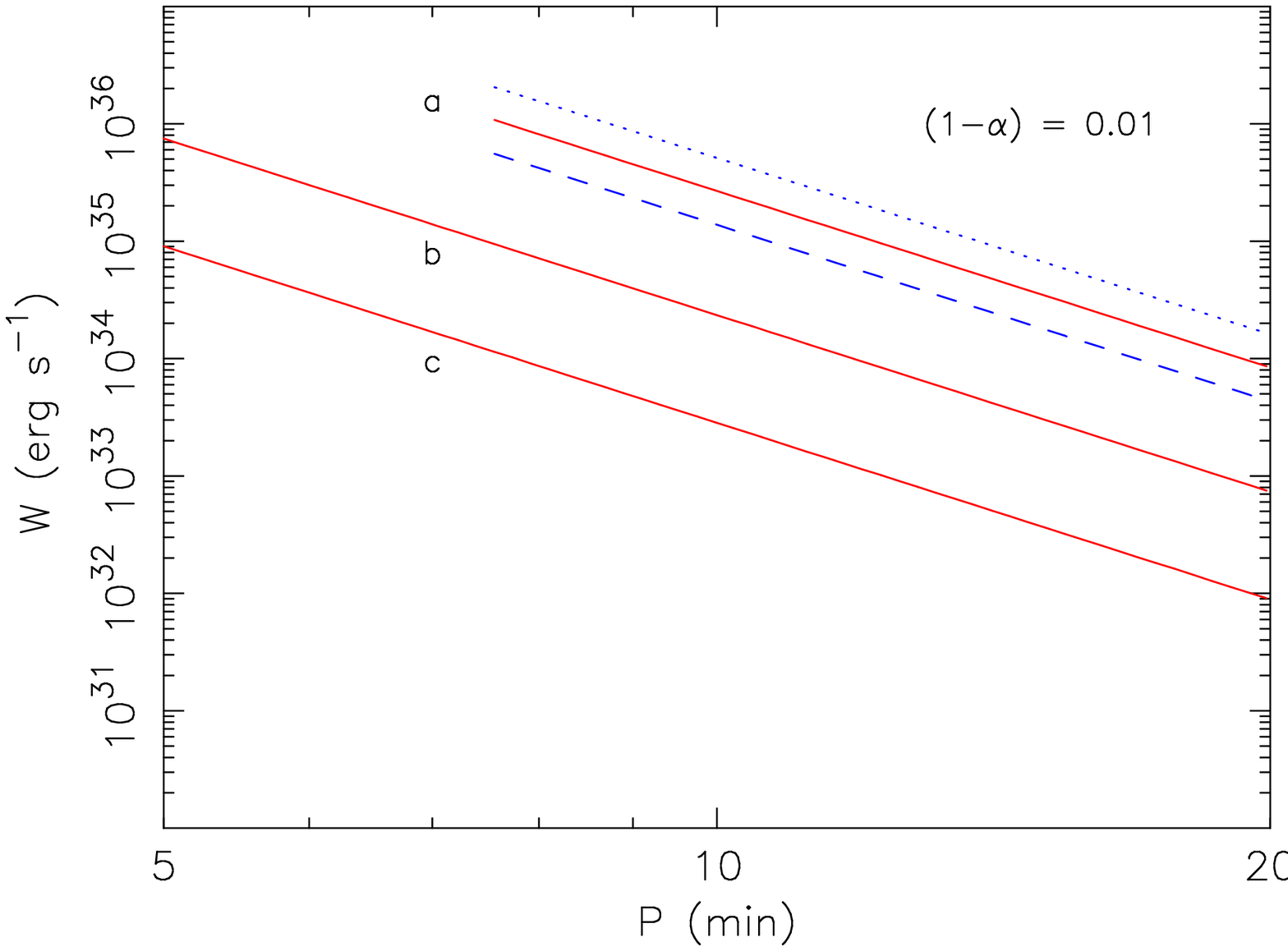} } 
   \caption{The total power generated by the dissipation of electric currents 
       as a function of the orbital period 
       for spin-orbit asynchronism $(1-\alpha)$ of 1/1000 and 1/100 
       (left and right panels respectively) predicted by the unipolar induction for UCDs.  
    The solid lines correspond to cases with a 1.0-M$_\odot$ magnetic white dwarf. 
    Lines a, b and c correspond to the cases with a non-magnetic companion white dwarf 
       of 0.1, 0.5 and 1.0~M$_\odot$ respectively. 
    The dotted line corresponds to the case with a 0.7-M$_\odot$ magnetic white dwarf  
          and a 0.1-M$_\odot$ non-magnetic white dwarf; 
      the dashed line,  a 1.3-M$_\odot$ magnetic white dwarf 
           and a 0.1-M$_\odot$ non-magnetic white dwarf.  
     The white-dwarf magnetic moments are $10^{32}$~G~cm$^{3}$ in all cases.  
     (Adapted from Wu, Ramsay \& Willes (2008).) 
     } 
   \label{ui_power} 
    \end{center} 
\end{figure}

The operation of a unipolar inductor in UCDs   
   can be understood in terms of an electric circuit model. 
The non-magnetic white dwarf, where the e.m.f. is generated, 
   acts as an electric generator or a  battery (with a small internal resistance); 
   the plasmas that mediate the currents are the conducting circuit wires; 
   and the magnetic white dwarf is the resistive load, where most of the dissipation occurs.  
The induced e.m.f.\ depends strongly on the binary orbital period, 
   the degree of spin-orbit synchronism,  and the mass (radius) of the non-magnetic white dwarf.     
The resistivities within the circuit, however, depend also 
  on the internal properties of the white-dwarf atmosphere.   
For a large range of mass ratios, 
   unipolar induction in a compact white-dwarf pair 
   can produce luminosities similar to or larger than the Sun, 
   requiring only a small degree of spin-orbit asynchronism  (Fig.~\ref{ui_power}). 

The remaining question now is: what actually drives the electric currents? 
The energy reservoir is in fact the binary orbit.  
Through unipolar induction, a back Lorentz torque is generated and it acts on the orbit.   
Orbital energy is extracted, which provides the e.m.f.\ for the current circuit.
Thus, similar to accretion, the ultimate energy source in a unipolar-inductor white-dwarf pair 
   is still the gravitational potential.  


\subsection{Candidate unipolar-inductor ultra-compact double degenerates}    

\begin{figure}
   \vspace{2mm}
   \begin{center}
   \hspace{3mm}\psfig{figure=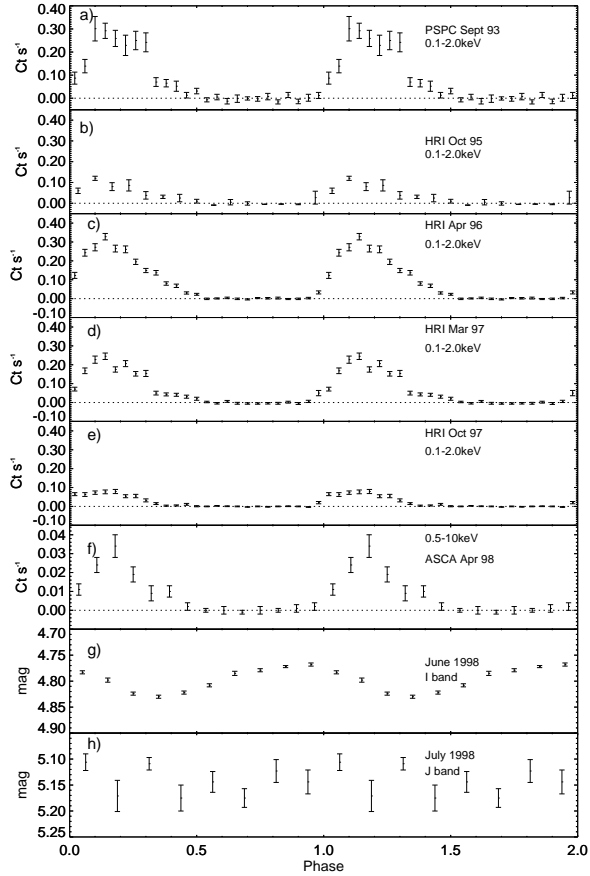,width=10cm,angle=0.0}
   \caption{
      The folded light curves of RX J1914+24 (Ramsay et al.\ 2000). 
      Panels (a) to (e) are X-ray light curves obtained by {\it ROSAT};  
          panel (f) is the X-ray light curve obtained by {\it ASCA}. 
      Panels (g) and (h) are the I and J band near-IR light curves 
          obtained by UKIRT.     
            }
   \label{rx1914_lightcurve}
   \end{center}
\end{figure}

\begin{figure}
   \vspace{0.02mm}
   \begin{center}
   \hspace{3mm}\psfig{figure=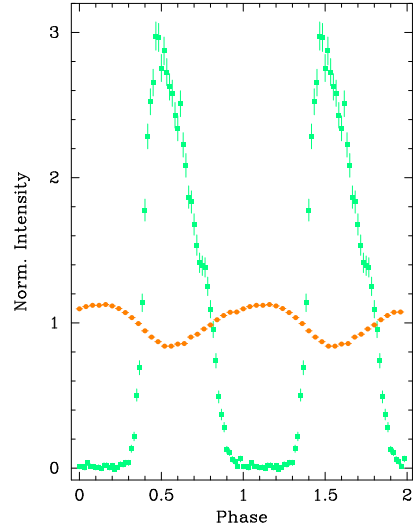,width=6.0cm,angle=0.0}
   \caption{The folded {\it Chandra} X-ray and VLT optical R-band light curve of 
     RX~J0806+15 (provided by G.~L.~Israel). 
            }
   \label{rx0806_lightcurve}
   \end{center}
\end{figure}

The two candidate unipolar-inductor UCDs, RX~J1914+24 and RX~J0806+15,  
  are short-period variable X-ray sources 
  discovered in the {\it ROSAT} observations 
  (Motch et al.\ 1996; Cropper et al.\ 1998; Israel et al.\ 1999). 
One of their remarkable characteristics 
  is that only a single period is shown in the variations across the electromagnetic spectrum 
  --- from the infra-red (IR) and optical  to X-ray bands 
  (see Fig.~\ref{rx1914_lightcurve} and \ref{rx0806_lightcurve}).   
The period of RX~J1914+24 is 569~s (Ramsay et al.\ 2002),  
   and the period of RX~J0806+15 is 321~s (Israel et al.\ 2003).  
Their X-ray light curves show pulse-like profiles, 
  suggesting that the emission originates from a hot spot  on the surface of one of the stars. 
The optical/IR light curves, in contrast, show sinusoidal variations, 
   and variations are anti-phased with the variations in the X-ray bands. 
The optical/IR emission region is therefore extensive 
   and not coincident with the X-ray emitting region.   


The nature of RX~J1914+24 and RX~J0806+15 has been under debate.  
It is now generally accepted that they are binary systems 
    with orbital periods of 569~s and 321~s respectively. 
This requires the two component stars in RX~J1914+24 and RX~J0806+15 
   to be degenerate stars.   
Moreover, they are very compact binaries with orbital separations similar to Jupiter's linear size.  
Several models for them been proposed: 
   (i) face-on intermediate polar (IP) (Norton, Haswell \& Wynn 2004), 
   (ii) degenerate polar (degenerate AM Herculis binary) (Cropper et al.\ 1998), 
   (iii) direct impact accretor (Marsh \& Steeghs 2002; Ramsay et al.\ 2002),   
   (iv) neutron star-white dwarf pair (Ramsay et al.\ 2002), 
   and (v) unipolar-inductor binary (Wu et al.\ 2002; Dall'Osso, Israel \& Stella 2006, 2007).   
In the first four models accretion is the energy source for  the observed X-rays.
The unipolar-inductor model, however, suggested that the emission of X-rays  
   is caused by ohmic dissipation of electric currents in the white-dwarf atmosphere. 
An assessment of the models  can be found in Cropper et al.\ (2004). 

In the face-on IP model, a moderately magnetized white dwarf is accreting material 
  from a main-sequence donor star. 
The white-dwarf spin is not synchronous with the orbital rotation. 
The pulse period of the X-ray emission is the white-dwarf spin period, 
  which is much shorter than the undetected orbital period.  
In the degenerate polar model, 
   the accreting white dwarf has a strong magnetic field.   
The mass-donor white dwarf may or may not be magnetic.    
The whole system is locked into synchronous rotation 
   by a white-dwarf magnetic field as in the usual polars (AM Herculis binaries).  
The observed period is the spin periods of the two white dwarfs.  
It is also the period of the orbital rotation.  
In the direct impact accretor model, both stars are white dwarfs. 
Their magnetic fields are irrelevant as they do not play a significant role 
  in determining the emission and the orbital dynamics.  
The X-ray hot spot is the stream impact point.  
Its location on the equator of the accreting white dwarf 
  is fixed in the rotational frame of the binary. 
The observed period is the orbital period and the spin period of the mass-donor white dwarf, 
  but it is not necessarily the spin period of the accreting white dwarf.   
In the neutron star-white dwarf pair model,   
   there is no mass transfer from the white dwarf to the neutron star. 
Otherwise, much higher X-ray luminosities would have been observed. 
There is only a low level of accretion, which is likely sustained 
  by remnant material in the vicinity of the binary   
  ejected in previous evolutionary phases.  

In the unipolar-inductor model, RX~J1914+24 and RX~J0806+15   
   contain one magnetic and one non-magnetic (or weakly magnetic) star. 
It allows the magnetic star to be a neutron star or a white dwarf, 
   but in a restrictive version both stars are white dwarfs.   
Electromagnetic radiation from these two systems 
  is not powered by accretion. 
Instead it is due to the dissipation of electric currents.  
The unipolar-inductor binary is in contrast to 
  other stellar objects whose energy sources are either accretion or nuclear reaction.    
A small asynchronism between the spin of the magnetic white dwarf and the orbital rotation 
  is required in order to generate a substantial e.m.f.\ which drives the electric currents. 
The focusing field lines channel the electric currents 
  toward a small foot-point region on the surface of the primary white dwarf.  
This gives a very small X-ray emission spot. 
The optical/IR emission is from a heated hemisphere of the secondary white 
  irradiated by the X-rays emitted from the primary white dwarf.  
The optical/IR emitting area is therefore extensive.  
This geometrical configuration naturally leads to an anti-phasing 
  between the optical/IR emission and X-rays.      

In order to account for the observed X-ray luminosity, 
  all accreting white-dwarf models 
  (the face-on IP, degenerate polar and direct-impact accretor model) 
  require a relatively high mass transfer rate.   
If we take the X-ray luminosity of $\sim 10^{35} - 10^{36}$~erg~s$^{-1}$ 
   (assuming a distance of 100~pc)  
   deduced for RX~J1914+24 from the {\it ROSAT} data (Cropper et al.\ 1998),  
   the mass transfer rate of the system exceeds $5 \times 10^{17}$~g~s$^{-1}$. 
Transfer of material from the low-mass secondary star to the high-mass primary star,  
   in general, causes the binary orbit to expand, 
   and hence the orbital period increases. 
For rapid mass transfer on timescales shorter than 
   the timescale of orbital evolution driven by angular momentum loss 
   (via gravitational radiation or magnetic braking),  
   the orbital period of the binary is expected to increase,  
   i.e.\  $\dot{P_{\rm o}}> 0$ (or ${\dot \omega}_{\rm o} < 0$).  
X-ray timing observations, however, show that the periods of these systems are decreasing 
  (Strohmayer 2002, 2003, 2004, 2005; Hakala et al.\ 2003; Ramsay et al.\ 2005), 
   which is inconsistent with the mass-transfer scenario. 
Accretion models are difficult to reconcile with the findings that  
   RX~J1914+24 has an almost featureless optical spectrum (Steeghs et al.\  2006) and  
   that RX~J0806+15 has only a few very weak optical emission lines (Israel et al.\ 2002).     
It is puzzling that signatures of accreting systems  
   such as the strong prominent H Balmer and He II emission lines 
   as those observed in cataclysmic variables (see Williams 1983) 
   and low-mass X-ray binaries (see Lewin, van Paradijs \& van den Heuvel 1997)  
   are not seen in the optical spectra of  RX~J1914+24 and RX~J0806+15. 
Recent {\it Chandra} observations of RX~J0806+15  
   confirmed that emission lines are absent in the X-ray band (Strohmayer 2008).   
Another difficulty of the scenarios with a strongly magnetic white dwarf  
   (as in the degenerate polar model) is non-detection of cyclotron harmonic features 
   in the optical spectra 
   (cf.\ the observed cyclotron humps in the optical spectra of AM Herculis binaries, 
   see e.g.\ Cropper et al.\ 1989).  
    
The unipolar-inductor model avoids the above difficulties of the accretion models.  
Although the model is generally consistent with existing observations 
  (see Cropper et al.\ 2004),   
  there are some concerns regarding whether or not it is applicable 
  to RX~J1914+24 and RX~J0806+15, which are presumably compact white-dwarf pairs    
  (e.g.\ Barros et al.\ 2005, 2007; Laine, Lin \& Dong 2008; Wood 2009).   
Some concerns, e.g. regarding the exact magnetic-field geometry  
  and the relative lead/lag in the X-ray pulses and the optical maxima, 
  can be resolved easily. 
There are, however, several more serious issues. 
For instance, certain implicit assumptions have been made  
  in order to facilitate the unipolar-induction process.   
In plasma, the time-dependence of a magnetic field is governed by 
\begin{eqnarray} 
  \frac{1}{c}  \frac{\partial {\vec B}}{\partial t} 
     & = & \nabla \times ({\vec \beta} \times {\vec B})  + \frac{c}{4\pi \sigma} \nabla^2 {\vec B} \ .  
\end{eqnarray}  
The first term on the right-hand side of the equation is the induction term, 
   and the second term is the diffusion term.  
In the unipolar-induction model, the diffusion term is omitted 
   based on the assumption that the white dwarf's core is a perfect conductor, 
   i.e., setting the conductivity $\sigma \rightarrow \infty$.   
This issue was discussed in detail recently by Laine, Lin \& Dong (2008) 
   in the context of binaries containing a normal star and a planet. 
The assumption is probably acceptable for compact white-dwarf pairs,  
   as the white-dwarfs' cores are practically a fermi ball of electrons.     
Another serious issue concerns the life-span of unipolar-inductor UCDs. 
If the system achieves spin-orbit synchronisation on a very short timescale,  
   the unipolar-induction process will be quenched.   
As the X-rays from unipolar-inductor UCDs are powered by electrical dissipation, 
   a rapid spin-orbit syncronisation would imply 
   that the X-ray active phases of the system are brief.  
If unipolar induction occurs only in transient episodes, 
   it would play a less important role 
  in determining the orbital evolution of UCDs.   
In \S~\ref{subsect:life_span}, we will discuss the operation of unipolar-induction, 
  spin-orbit synchronisation and life-span of unipolar induction UCDs in more detail.   
  

\section{Orbital evolution of unipolar-inductor compact binaries} 
\label{sect:orbital_de}

\subsection{Spin-orbit coupling} 
\label{subsect:spin_coupling}

In the unipolar inductor model, the system is asynchronous,  
  and the orbital evolution is described neither by 
  Equation~(\ref{eq:orbit_1}) nor (\ref{eq:orbit_2}) in Section~\ref{sect:dynamics}. 
Additional energy dissipation needs to be taken into account. 
Without loss of generality, 
   we consider the non-magnetic white dwarf as tidally locked 
   to synchronous rotation with the orbit. 
This is justified if the secondary white dwarf is close to filling its Roche lobe.   
Through spin-orbit coupling, energy and angular momentum 
  are transferred between the binary orbit and the spin of the magnetic white dwarf,  
  but the transfer rates depend on the orbital properties and the dissipation processes. 

We may define a quantity  
 \begin{eqnarray} 
   W^{*}  & \equiv  & \frac{W}{\left(1-\alpha \right)^{2}} \ .  
\label{eq:w_star}
\end{eqnarray}
This quantity is independent of the asynchronism parameter $\alpha$,  
   and it  gives  the timescale on which the system achieves spin-orbit synchronism.   
The two essential equations governing the spin-orbit  evolution are 
\begin{eqnarray}
  \frac{\dot{\omega}_{\rm o}}{\omega_{\rm o}} & = & \frac{{\dot E}_{\rm gw}}{g(\omega_{\rm o})}
  \left[ 1 -  (1-\alpha)  \frac{W^*}{{\dot E}_{\rm gw}}\right] \  , 
\label{eq:omega_evol} \\ 
  \frac{\dot{\alpha}}{\alpha} & = & - \frac{\dot{E}_{\rm gw}}{g(\omega_{\rm o})}
  \left[ 1  - (1-\alpha) 
  \left( 1 + \frac{g(\omega_{\rm o})}{\alpha I_{1} \omega_{\rm o}^{2}} \right) 
   \frac{W^*}{\dot{E}_{\rm gw}}   \right]    
\label{eq:alpha_evol}
\end{eqnarray} 
   (Wu et al.\ 2002), where  
\begin{eqnarray}
   g(\omega_{\rm o}) &  =  & - \frac{1}{3}
    \left[ \frac{q^{3}}{1+q} G^{2} M_{1}^{5} \omega_{\rm o}^{2} \right]^{1/3} \, 
    \left[1 - \frac{6}{5}  (1+q) f(\omega_{\rm o}) \right]  \nonumber \\ 
    & = & - \frac{1}{3} G^{2/3} M_{\rm chirp}^{5/3} \omega_{\rm o}^{2/3} 
      \left[1 - \frac{6}{5}  (1+q) f(\omega_{\rm o}) \right]  \ . 
\label{eq:g_factor}
\end{eqnarray}  
The structure factor $f(\omega_{\rm o})$ is  
\begin{eqnarray}
f(\omega_{\rm o}) & = & \left[ 
    \frac{R_{2}^{3} \omega_{\rm o}^{2}}{G (M_{1} + M_{2})} \right]^{2/3}  \ .  
\label{eq:f_factor}
\end{eqnarray} 
The moment of inertia of the magnetic white dwarf, 
    $ I_1 =   {2}\eta M_{1} R_{1}^{2}/ 5$, 
  and the parameter $\eta$ depends on the density distribution and shape of the white dwarf.  
For spherical stars with a uniform density, $\eta = 1$. 
Note that by setting $W^* = 0$ and considering $\lim f(\omega_{\rm o}) \rightarrow 0$, 
    we can recover the expression of $\omega_{\rm o}/\omega$ 
    for the case with no spin-orbit coupling (Eq.~\ref{eq:orbit_1}).  
    
\subsection{Life span of unipolar-inductor compact binaries}
\label{subsect:life_span}

The right hand side of Equation~(\ref{eq:alpha_evol}) is dominated by the final term in the bracket.  
The synchronisation of the system due to electrical dissipation 
  in the unipolar-inductor circuit is essentially governed by the equation 
\begin{eqnarray}
  \alpha & \approx &  1- (1- \alpha_{0})~ \exp \left[ - \frac{t}{\tau_{\rm ui}} \right] \ , 
\label{eq:alpha_approx}
\end{eqnarray}
  where 
\begin{eqnarray}
  \tau_{\rm ui} & =  & \frac{I_{1} \, \omega_{\rm o}^{2}}{W^{*}}  
\end{eqnarray} 
   is the synchronisation (unipolar-induction) timescale. 
An approximate expression for $\alpha_0$ as a function of $M_1$ and $M_{2}$  
  can be obtained by solving Equation~(\ref{eq:omega_evol}) for $\alpha$, with
\begin{equation}
\alpha_{0} \sim  1 - \frac{1}{W^{*}} \left[ \dot{E}_{\rm gw} - 
  g(\omega_{\rm o}) \frac{{\dot \omega}_{\rm o}}{\omega_{\rm o}} \right]  
\end{equation}
   (Willes \& Wu, unpublished). 
The lifetime of the system is limited  by gravitational radiation loss, 
  with the merging timescale for the binary system given by 
\begin{equation}
\tau_{\rm gw} = \frac{a_{0}^{4}}{4\ \Theta}   \  , 
\end{equation}
   where $a_{0}$ is the initial binary orbital separation, and
\begin{eqnarray}
\Theta & = &    a^3 {\dot a}     \\ 
   & = &   \frac{64}{5} \frac{G^{3}}{c^5} \left [ M_{1} M_{2} (M_{1} + M_{2}) \right]   
\end{eqnarray}
   (Peters 1964; see also Landau \& Liftshitz 2002).  

The synchronisation timescale appears to be very short, with $\tau_{\rm ui} < 1000$ yr, 
  for a range of combinations of white-dwarf mass ($M_{1}$, $M_{2}$) and magnetic moment ($\mu$). 
One might be concerned  
   that unipolar-induction systems can be X-ray sources 
   over only a small fraction of the binary system lifetime $\tau_{\rm gw}$,
   which is the timescale for white-dwarf coalescence due to gravitational radiation losses.
The apparently short X-ray emission phase could pose problems 
   for the detectability of these systems, as pointed out by Barros et al.\ (2005).
A possible resolution is to invoke a mechanism, 
   such as intermittent mass transfer, 
   which causes repeated episodes of spin-orbit de-synchronisation over the system lifetime.  
While this is possible, it is not always necessary.   
In the parameter regimes of UCDs, 
  the unipolar induction phase can operate and produce X-ray pulses 
  over the system lifetime before coalescence occurs, 
  even when $\tau_{\rm ui} \ll \tau_{\rm gw}$.    

  
\begin{figure}
   \vspace{2mm}
   \begin{center}
   \hspace*{8mm}
   \psfig{figure=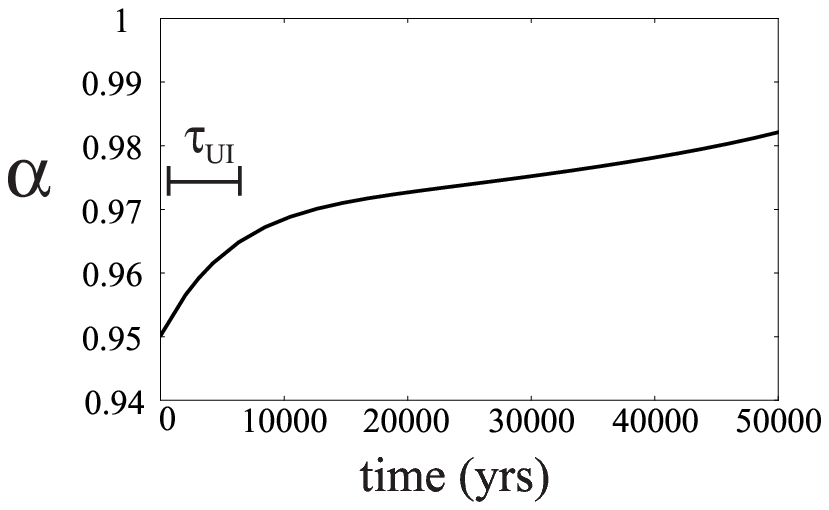,width=9cm,angle=0.0} 
    \hspace*{-9mm} 
   \psfig{figure=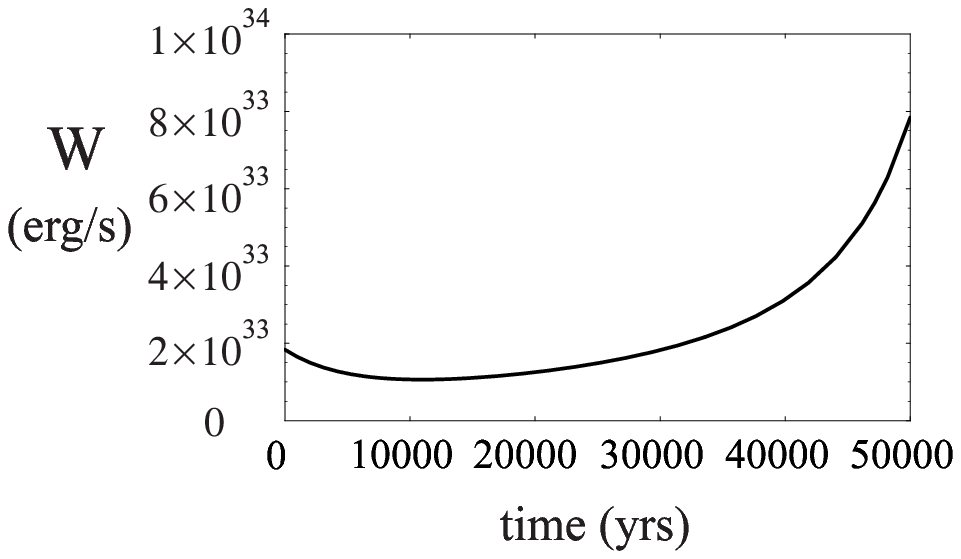,width=9.75cm,angle=0.0}
   \caption{
       (Top) Evolution of the asynchronicity parameter $\alpha$ for a compact white-dwarf pair  
             with $M_1 = 0.7\ {\rm M}_{\odot}$ and $M_2 = 0.345\ {\rm M}_{\odot}$
             and $\mu = 10^{30}  \, {\rm G} \, {\rm cm}^{3}$. 
       The orbital period and its time derivative 
             take the values derived for RX~J0806+15 (see  Israel et al.\ 2003; Hakala et al.\ 2003). 
      The evolution is characterized by a rapid phase (up to $t \sim \tau_{\rm ui}$) 
             followed by a slow phase 
             where the evolution is controlled the dynamics of the orbital decay 
             associated with gravitational radiation losses. 
      For this set of parameters, 
             the system would not be completely synchronised when entering the slow phase. 
     (Bottom) Evolution of the footpoint luminosity (electrical dissipation) $W$ 
             associated with $\alpha$ from the top panel. 
     The value of $W$ remains above the level of $\sim 10^{33} \, {\rm erg}~{\rm s}^{-1}$ 
             throughout the entire binary-system lifetime, 
          and at the later evolutionary stage it even increases 
             despite the spin and orbit becoming more synchronised.  
      (Diagrams provided by A.~Willes.)
            }
 \label{orbital_evolution} 
   \end{center} 
\end{figure} 


This phenomenon can be illustrated with the following example. 
Consider a system with white-dwarf masses 
  $M_1 = 0.7\ {\rm M}_{\odot}$ and $M_2 = 0.345\ {\rm M}_{\odot}$
  and with the primary white dwarf having a magnetic moment $\mu = 10^{30} \, {\rm G} \, {\rm cm}^{3}$.
For these parameters,  the initial value of the asynchronism parameter   
  required to fit the observed orbital period $P_{\rm o}$ and period derivative ${\dot P}_{\rm o}$ 
  for RX~J0806+15 is $\alpha_{0} = 0.95$.   
Figure~\ref{orbital_evolution} shows the evolution of $\alpha$ over a 50000 year period.  
The synchronisation timescale of the system $\tau_{\rm ui} \sim 5000$~yr.
After a brief unipolar-inductor phase ($t > \tau_{\rm ui}$), 
  the driver of system evolution is taken over by the gravitational radiation loss. 
However, the system has not achieved synchronism  
  by the end of the unipolar-inductor phase ($\alpha \ne 1$).  
During the subsequent slow evolution, 
  gravitational radiation loss ensures 
  that the system remains asynchronous over the remaining time span 
  until the eventual coalescence of the two white dwarfs.   
Note that the rate of change of $\alpha$ is effectively zero 
  (in comparison to the fast evolution of $\alpha$ during the unipolar-inductor phase), 
  despite the fact that the system is in an asynchronous state.
The value of $\alpha$ at the ``end'' of the unipolar-inductor phase 
  can be estimated by equating the first and last terms 
  on the right-hand side of Equation~\ref{eq:alpha_evol}  
  (i.e.\  by setting $\dot{\alpha}=0$, and 
  where $g(\omega_{\rm o})/(\alpha I_{1} \omega_{\rm o}^{2}) \gg 1$), 
  yielding
\begin{eqnarray}
   \alpha_{\rm gw} & = &  \frac{\chi}{1+\chi} \  , 
\end{eqnarray}
  where
\begin{eqnarray}
   \chi & = & \frac{g(\omega_{\rm o})}{\dot{E}_{\rm gw} \, \tau_{\rm ui}} \ .
\end{eqnarray}
For these parameters, $\alpha_{\rm gw} = 0.97$, 
  which is in approximate agreement with the value of $\alpha$ 
  at the end of the unipolar-inductor  phase, $t \sim 5000$~yr  
  (top panel, Fig.~\ref{orbital_evolution}).
The system remains unsynchronised, with $\alpha \approx 0.98$,   
  over a period of $t  \sim 50000$~yr. 
Throughout the evolution, 
  the footpoint luminosity (electrical dissipation) $W$ 
  exceeds $10^{33} \, {\rm erg}~{\rm s}^{-1}$ (bottom panel, Fig.~\ref{orbital_evolution}). 

We note that for some parameters,  
   a system can achieve a high degree of synchronisation on a short timescale. 
For instance if we consider different masses for the white dwarf, 
   say $M_1 = 0.7\ {\rm M}_{\odot}$ and $M_2 = 0.1\ {\rm M}_{\odot}$, 
   then the system is almost completely synchronised 
   and $\alpha_{\rm gw}$ reaches  0.998 within 1500 yr, 
   and the corresponding footpoint luminosity $W$ falls 
   below  $10^{32} \, {\rm erg}~{\rm s}^{-1}$ soon after this time.
Here we have  demonstrated that over a certain range of parameters, 
  the unipolar-inductor model can sustain intense X-ray emissions over the entire lifetime of the system, 
  rather than the much shorter unipolar-inductor timescale.  
  
Compact white-dwarf pairs are strong sources of gravitational radiation. 
They are populous in the solar neighbourhood and are among the first 
    to be detected by the gravitational wave observatory {\it LISA}   
    (see Cutler, Hiscock \& Larson 2003; Nelemans 2003; 
    Nelemans, Yugelson \& Portegies Zwart 2004; Kopparapu \& Tohline 2007).  
These sources can be calibrators of the experiments 
  or pests that cause foreground contamination of the weaker cosmological signals.  
In order to have these sources detected and subtracted,     
  one needs good waveform templates of the gravitational radiation that they emit. 
As illustrated above, the unipolar-induction can persist throughout the lifetime of a UCD 
   until the two white dwarfs coalesce.  
The orbital evolution is determined by the energy loss due to gravitational radiation  
   and electrical dissipation. 
Without taking into account the contribution of unipolar induction,  
   the gravitational wave signals of a UCD can become de-coherent 
   on timescales as short as days, 
   thus posing serious problems in the UCD detection. 
  

\begin{figure}
   \vspace{5mm}
   \begin{center}
   \hspace{3mm}\psfig{figure=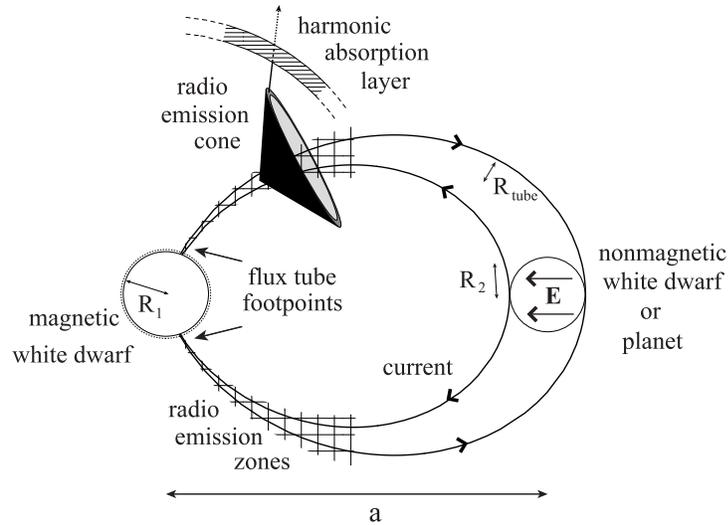,width=9.5cm,angle=0.0}
   \caption{
      An illustration to show the regions where electron-cyclotron masers would emit 
          in a compact white-dwarf pair or a white-dwarf planetary system. 
            }
   \label{maser_geometry}
   \end{center}
\end{figure} 


  
\begin{figure}
   \vspace{2mm}
   \begin{center}
   \hspace{0mm}
   \psfig{figure=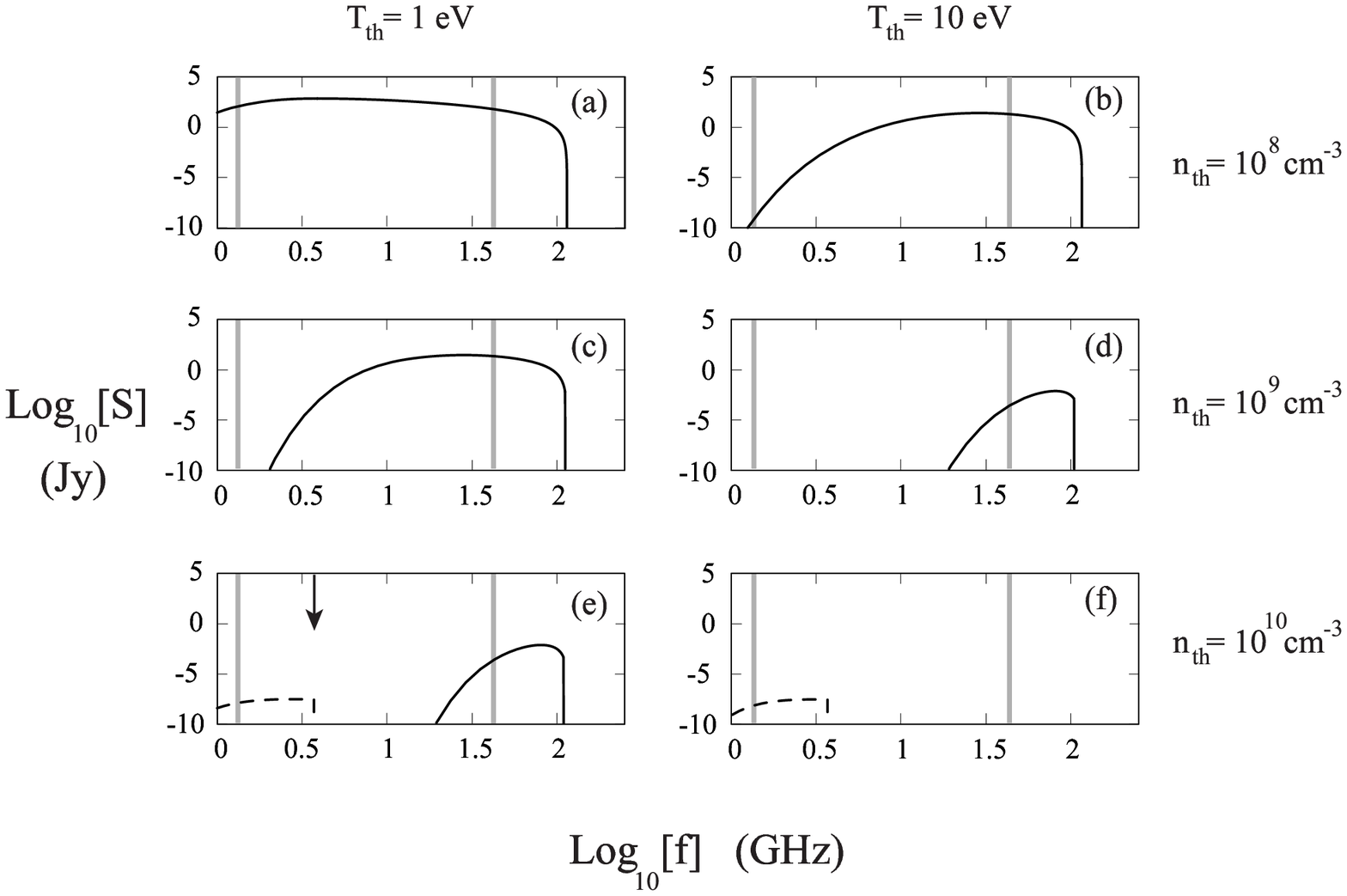,width=13.2cm,angle=0.0}
   \caption{
       Peak flux densities of electron-cyclotron maser (in the radio wave-bands) 
            from unipolar-inductor UCDs (maximised over emission angle).   
       The system parameters of the UCD are 
            white-dwarf masses $M_1 = 0.7~{\rm M}_\odot$ and $M_2 = 0.5~{\rm M}_\odot$, 
            orbital period $P_{\rm o} = 540~{\rm s}$, 
            magnetic moment of the primary white dwarf $\mu = 10^{31}~{\rm G}~{\rm cm}^3$ 
            and degree of asynchronism of 1 part in 1000. 
       The loss-cone parameters are 
            the temperature $kT = 1~{\rm keV}$, 
            the electron number density $n_{\rm lc} = 10^9~{\rm cm}^{-3}$, 
            and the edge width $\Delta \alpha = 0.05$ (see Willes \& Wu 2004). 
       Left and right columns correspond to cases with thermal electrons of temperatures 
           $kT_{\rm th} =$ 1 and 10~eV respectively. 
       Panels from top to bottom correspond to thermal electron number density  
           $n_{\rm th} =  10^8$, $10^9$ and $10^{10}~{\rm cm}^{-3}$ respectively. 
       The x-mode emission is represented by solid lines, 
           and the o-mode emission by dashed lines. 
       The vertical grey lines mark the VLA observing frequencies 
           of 1.465 and 43 GHz.       
                }
   \label{maser_wd2}
   \end{center}
\end{figure}  



\section{Electron-cyclotron maser emission}   
\label{sect:maser}

UCDs are potential electron-cyclotron maser sources.       
The two distinguishable characteristics of electron-cyclotron masers 
   are high brightness temperature and almost 100\% circular polarization.  
The operation of electron-cyclotron masers requires 
  a population inversion in the electron distribution
  and  a magnetised plasma in which
  the electron-cyclotron frequency $\Omega_{\rm e}$
  exceeds the plasma frequency $\omega_{\rm p}$  (e.g.\ Dulk 1985). 
These two conditions can be satisfied in a variety of astronomical settings.     
The first condition can be achieved in the presence of a loss-cone or a shell electron distribution. 
These distributions are kinetically unstable, 
  and the instability provides the free energy for the generation of electron-cyclotron masers  
 (Wu \& Lee 1979; Melrose \& Dulk 1982; Pritchett 1984; Melrose 2005; Treumann 2006)  
A loss-cone electron distribution arises  
  when an electron pitch-angle anisotropy develops 
  within a magnetic flux tube with converging field lines at each foot point. 
Large pitch angle electrons are magnetically reflected, 
  whereas small-pitch-angle electrons are lost through collisions 
  with high density plasma at the foot of the magnetic flux tube.
The second condition is satisfied
  in magnetized plasmas with a relatively low electron density
  and/or a high magnetic field strength. 

In Jupiter and Io, electron cyclotron masers are emitted  
  from the current-carrying electrons in the Io magnetic flux tube.  
The observed high brightness temperatures ($\ga 10^{17}$~K, Dulk 1970), 
  $100\%$ circularly polarization (Dulk, Lecacheux \& Leblanc 1992) 
  and the radiation beaming pattern in the radio emission from Jupiter-Io 
  are characteristics of electron-cyclotron masers.    
The anti-correlation 
  between infrared footpoint emission and Io-controlled Jovian decametric radiation 
  indicates that the masers are driven by reflected electrons (Connerney et al.\ 1993). 
The presence of reflected electrons in a loss-cone distribution is also consistent 
  with the observation of negative frequency drifts 
  in the fine-frequency structure of Jovian decametric radiation S-bursts (Ellis 1974).  
    
The operation of unipolar induction  
  and the similarity of configurations between a unipolar inductor UCD and the Jupiter-Io system 
  imply that loss-cone instability may develop 
  in the magnetic flux tubes in a UCD (see Fig.~\ref{maser_geometry}).  
The main differences between a UCD and the Jupiter-Io system  
  are probably the energetics of the streaming electrons in the current circuits, 
  which participate in developing the loss-cone instability,  
  and the amount of thermal electrons present in the system, 
  which could suppress the maser process. 
A model of electron-cyclotron masers from white-dwarf pairs 
  can be constructed in the unipolar-inductor framework 
  (see Willes \& Wu (2004) and Willes, Wu \& Kuncic (2004) for details).   
The predicted flux densities of electron-cyclotron masers from UCDs  
  with parameters are shown in Figure~\ref{maser_wd2}.      
For parameters similar to those derived for RX J1914+24 and RX J0806+15,  
   the electron-cyclotron masers are observable using current instruments  
   such as the radio telescopes ATCA and VLA.   
A radio survey would identify unipolar-inductor UCDs  
   which emit only weak X-rays  or have a very soft X-ray spectrum.  

Note that a recent search for electron-cyclotron masers from UCDs  
  (Ramsay et al.\ 2007) revealed a 5-$\sigma$ source at the position of RX~J0806+15. 
The inferred brightness temperature exceeded $10^{18}$~K 
  and the upper limit for circular polarization was about 50{\%}.  


\begin{figure}
   \vspace{5mm}
   \begin{center}
   \hspace{3mm}\psfig{figure=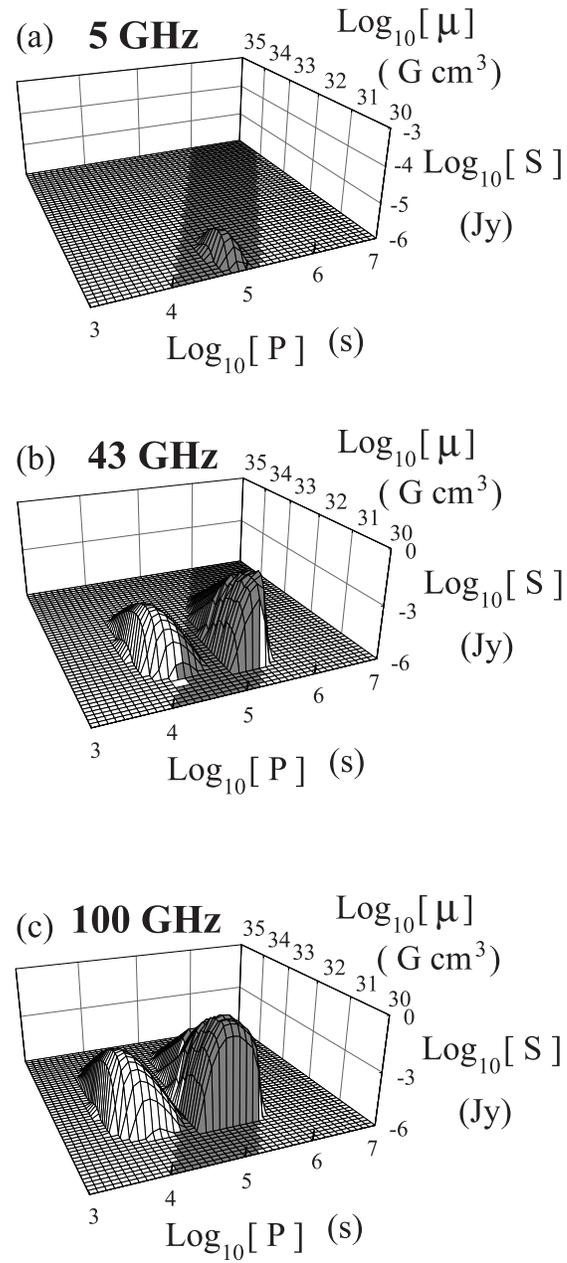,width=7.5cm,angle=0.0}
   \caption{ The flux density of loss-cone electron-cyclotron masers at 5, 43 and 100~GHz 
            from a UI white-dwarf planetary system. 
            The magnetic moment of the magnetic white dwarf $\mu = 10^{30}$~G~cm$^3$, 
       the number density of the loss-cone electron population is  $10^7$~cm$^{-3}$,  
       and the mean energy of the electrons is 1~keV.    
         The distance to the source is 100~pc. 
        The gray band denotes the region in the parameter space
            where detection is expected for the current instrumentation.        }
   \label{maser_wd-p}
   \end{center}
\end{figure} 



\section{Beyond ultra-compact double degenerates}  
\label{sect:beyond_ucd}

\subsection{White-dwarf planetary systems} 
\label{subsect:wd_planet}
     
Electron-cyclotron masers can also be generated in astronomical binaries 
   having similar configurations for electric currents and magnetic fields as a unipolar-inductor UCD. 
An example is a magnetic white dwarf with an orbiting terrestrial planet with a metallic core  
   (Li, Ferrario \& Wickramasinghe 1998). 
The metallic core is a good electric conductor. 
It provides the e.m.f.\ to drive the current flow.  
A unipolar-inductor white-dwarf planetary system differs from a unipolar-inductor UCD 
  by replacing the non-magnetic white dwarf with a less massive terrestrial planet.    
Electron-cyclotron maser generation 
   is determined by the magnetic field and electric current configurations, 
   and the charge acceleration processes.  
The strength of the masers has a very strong dependence  
   on the system size but a weak dependence on the system mass. 
Because a terrestrial planet and a white dwarf have similar sizes,    
    electron-cyclotron masers in white-dwarf planetary systems  
   can be as strong as in UCD (Willes \& Wu 2004, 2005).   
The predicted flux densities of loss-cone electron-cyclotron masers 
  from a unipolar-inductor white-dwarf planetary system 
  can exceed 0.1~Jy  for certain sensible system parameters 
  (see Fig.~\ref{maser_wd-p}).   

Can white-dwarf planetary systems be formed? 
White dwarfs are remnants of solar-like and low-mass stars. 
After evolving beyond the main-sequence and the giant phases,  
  the sun will become a white dwarf. 
For a solar-like system, 
  if the inner planets can survive 
  being engulfed by the inflated stellar envelope during the red-giant/asymptotic-giant phases, 
  the system will become a white-dwarf planetary system 
  (see discussions in Willes \& Wu 2005).  
Provided that the terrestrial planets spiral in sufficiently close to the white dwarf 
  so that efficient unipolar induction can operate, 
  a loss-cone or shell electron distribution may develop,  
 leading to the emission of electron-cyclotron masers.         
A population synthesis (Willes \& Wu 2005) 
   suggested that 
   there would be about five systems expected to be detected by VLA at 43~GHz,  
   about 20 systems by SKA at 20~GHz,  and 
   about 100 systems by ALMA at 100~GHz.     

\subsection{Einstein-Laub effect in compact binaries} 
\label{subsect:einstein-laub}  
  
Einstein and Laub (1908) pointed out that 
   a magnetic dipole moment moving in a constant velocity 
   would develop an electric dipole moment, i.e.\   
\begin{eqnarray} 
  {\vec d} & = & {\vec \beta} \times {\vec \mu}   \ , 
\end{eqnarray}  
 where ${\vec E}$ is the electric field, ${\vec B}$ the magnetic induction,  
   ${\vec \beta}$ the velocity normaised to the speed of light in vacuum, 
   ${\vec d}$ the electric dipole moment, and ${\vec \mu}$ the magnetic dipole moment.   
This essentially says that electrodynamics is a restrictive case of special relativity.        
However, it is not easy to set up an experiment 
  in which a strong magnet moves at a relativistic speed 
  so as to induce a measurable electric dipole moment.  
 
In a slab of insulating material moving with a constant velocity,  
   we would expect an electric polarisation $P$, given by 
\begin{eqnarray} 
  {\vec P} & = & \frac{\epsilon -1 }{4 \pi} \left( {\vec E} + {\vec \beta} \times {\vec B}\right) 
         + \left( {\vec \beta} \times {\vec m}  \right) \ , 
 \end{eqnarray} 
  where ${\vec m}$ is the magnetic polarisation and $\epsilon$ is the dielectric constant.  
Wilson and Wilson (1913) conducted a rotating-cylinder experiment, 
  which appeared to have verified this effect. 
However, the interpretation of their results and 
   whether the experiment is a validation of the effect has been under debate  
  (see Pellegrini \& Swift 1995; Weber 1997; Krotkov et al.\ 1999; Hertzberg et al.\ 2001).   
The arguments centre on the fact that 
  a spinning device was used in the Wilson \& Wilson experiment 
  while rotation is not equivalent to translational motion. 

The Einstein-Laub effect was subsequently verified in a molecular beam experiment 
 (Sangster et al.\ 1993; 1995), 
  which was designed for other scientific objectives.    
In the experiment,  
   a beam of magnetically polarised thallium fluoride molecules 
   (the magnetic dipoles with moments ${\vec  \mu}$) with a velocity ${\vec \beta}$ 
   was sent through a region of constant electric field ${\vec E}$. 
By measuring the Ahronov-Casher phase shift, which is given by 
\begin{eqnarray}  
  \hbar  \phi_{ab} & = & 
      \int^b_a dt \left( {\vec \mu} \times {\vec E} \right) \cdot {\vec \beta}  \nonumber \\ 
      & = &  \int^b_a dt \left( {\vec \beta} \times {\vec \mu} \right) \cdot {\vec E}  \nonumber \\ 
      & = &  \int^b_a dt ~{\vec d} \cdot {\vec E}   \   , 
\end{eqnarray} 
   where $\hbar$ is the reduced Planck constant. 
Sangster et al.\ (1995) deduced the interaction energy 
   and used it to infer the induced dipole moment $\vec d$.  
The result agreed with the theoretical prediction by relativity to within 2\%.  

We note that the Einstein-Laub effect in a rotating device 
   is manifested in compact binaries. 
Consider a magnetised white dwarf 
   rotating in a tight orbit around another compact object.   
The white dwarf has a magnetic moment ${\vec \mu}$. 
For simplicity, the magnetic moment ${\vec \mu}$ is  perpendicular 
   to the orbital angular velocity ${\vec \omega}_{\rm o}$.  
Moreover, the white-dwarf spin is magnetically locked 
   into synchronous rotation with the orbit, as in the AM Herculis binaries. 
The orbital rotation of the white dwarf would then induce a spinning electric dipole moment 
   with a magnitude 
\begin{eqnarray} 
   d & = & \frac{\mu r_{\rm o}\omega_{\rm o} }{c}  \ , 
\end{eqnarray}  
  where $r_{\rm o}$ is the radius of the white-dwarf orbit  
  with respect to the centre of mass of the binary.   
A spinning electric dipole is known to emit electromagnetic waves, 
The radiative power is given by  
\begin{eqnarray}  
  L  &  =  &  \frac{2}{3} \frac{\ddot{d}^2}{c^3}  
      = \frac{2}{3} \frac{\mu^2 r^2_{\rm o} \omega_{\rm o}^6}{c^5} \ . 
\end{eqnarray}    
For a binary with  $P_{\rm o} \sim 300$~s, 
   $r_{\rm o} \sim 10^{10}$~cm, 
   and a white dwarf with $\mu \sim 10^{33}$~G~cm$^3$, 
   the radiative power will be $L \sim 2.3 \times 10^{23}$~erg~s$^{-1}$.  
This value is similar to that of thermal emission from a spherical body 
  with a temperature of about 300~K and an Earth-sized radius. 
However, $r_{\rm o} \propto a_{\rm o} \propto \omega_{\rm o}^{-2/3}$, 
  implying that $L\propto \omega_{\rm o}^{14/3}$ 
   (cf.\ $L_{\rm gw} = {\dot E}_{\rm gw}  \propto \omega_{\rm o}^{10/3}$ for gravitational radiation).   
For a system with $P_{\rm o} \sim 5$~s (possible for two neutron stars in a merging process), 
   the expected radiative power would exceed $10^{31}$~erg~s$^{-1}$, 
   which would have some observational consequences.   
   

\section{Summary}   
\label{sect:summary}

Ultra-compact double degenerates contain two compact stars 
  revolving around each other in a very tight orbit.  
The proximity of the two stars allows efficient magnetic coupling 
  between the stellar spins and the orbital rotation.  
The presence of unipolar induction in compact binaries 
  could greatly affect the orbital dynamics in compact binaries, 
  leading to observational consequences  
  in gravitational radiation as well as in electromagnetic radiation domains. 
Unipolar-inductor compact binaries are possible strong sources 
  of electron-cyclotron masers. 
The maser model for unipolar-inductor ultra-compact double degenerate 
  can be applied to white-dwarf planetary systems. 
Einstein-Laub effects may be observable in compact binaries with extremely short orbital periods.  

\begin{acknowledgements} 
I thank Jingxiu Wang for the suggestion to write this review 
  and continuous encouragement throughout the writing process. 
Andrew Willes, Mark Cropper and Gavin Ramsay have been my collaborators 
  in various research projects on compact binary systems.  
They have contributed much to the science discussed in this article.  
I thank Andrew Willes for providing results of his unpublished work  
  on the evolutionary history of unipolar-inductor UCDs 
  and, in particular, Figure~\ref{orbital_evolution}. 
I also thank Gian-Luca Israel for providing Figure~\ref{rx0806_lightcurve},  
  and Ziri Younsi and Curtis Saxton for reading through the manuscript.     
\end{acknowledgements}

\label{lastpage}

\end{document}